\documentclass[lettersize,journal]{IEEEtran}

\usepackage{amsmath,amsfonts, amssymb}
\usepackage{comment}
\usepackage{algorithm}
\usepackage{algpseudocode}
\usepackage{array}
\usepackage{graphicx}
\usepackage{booktabs}
\usepackage{tabularx}
\usepackage{url}
\usepackage[noadjust]{cite}

\usepackage{orcidlink}
\usepackage{tikz}
\usetikzlibrary{positioning, shadows}
\usetikzlibrary{shapes.geometric, arrows.meta}
\usetikzlibrary{shapes,arrows}
\usetikzlibrary{decorations.pathmorphing,decorations.markings,calc}

\definecolor{comsocblue}{HTML}{005F9E}
\usepackage[]{hyperref}
\hypersetup{
    pdfstartview={FitH},
    colorlinks=true,
    linkcolor=black,
    citecolor=comsocblue,
    urlcolor=comsocblue,
}
\usepackage{cleveref}
\crefname{figure}{Fig.}{Figs.}

\usepackage{cleveref}
\usepackage[caption=false,font=footnotesize]{subfig}

\crefname{figure}{Fig.}{Figs.}
\Crefname{figure}{Fig.}{Figs.}

\usetikzlibrary{fit}
\pgfdeclarelayer{bg}
\pgfsetlayers{bg,main}

\usepackage{acronym}
\usepackage{doi}

\usepackage{pgfplots}
\pgfplotsset{compat=1.18}
\usepackage[dvipsnames, svgnames]{xcolor} 

\usepackage{setspace}
\usepackage{float}

\newcommand{\Tr}{\operatorname{Tr}}
\newcommand{\E}{\mathcal{E}}
\newacro{FSO}{free-space optical}
\newacro{RF}{radio frequency}
\newacro{UAV}{unmanned aerial vehicle}
\newacro{SLIPT}{simultaneous lightwave information and power transfer}
\newacro{VLC}{visible light communication}
\newacro{OWC}{optical wireless communication}
\newacro{LED}{light-emitting diode}
\newacro{QoS}{quality of service}
\newacro{TS}{time switching}
\newacro{PS}{power splitting}
\newacro{BER}{bit error rate}
\newacro{DC}{direct current}
\newacro{AC}{alternating current}
\newacro{EH}{energy harvesting}
\newacro{IM/DD}{intensity modulation and direct detection}
\newacro{OOK}{on-off keying}
\newacro{SNR}{signal-to-noise ratio}
\newacro{NOMA}{non-orthogonal multiple access}
\newacro{RIS}{reconfigurable intelligent surface}
\newacro{IRS}{intelligent reflecting surface}
\newacro{ID}{information detection}
\newacro{CW}{continuous-wave}
\begin{document}
\title{QCommE2E: An Open-Source Simulation of End-to-End Quantum Communication Systems}

\author{Omar Alnaseri \orcidlink{0000-0002-7246-3974},~\IEEEmembership{Senior Member,~IEEE}

\thanks{Omar Alnaseri is with the Department of Electrical Engineering, Baden-Wuerttemberg Cooperative State University, Friedrichshafen, Germany (Corresponding author: Omar Alnaseri, email: alnaseri.omar@dozent.dhbw-ravensburg.de).}
}

\maketitle

\begin{abstract}
This paper presents \textit{QCommE2E} as an open-source simulation framework for end-to-end quantum communication systems, with explicit tutorial emphasis. The primary objective is to develop a comprehensive framework that includes transmitters, receivers, communication channels, performance metrics, and visualization tools, to facilitate the systematic design, configuration, and analysis of experimental simulations for novel quantum communication architectures. As the primary use case, we walk through the current quantum channel comparison, which maps textbook quantum-information channels and reduced optical-fiber/free-space surrogates into a single executable benchmark. We describe the common density-matrix interface, the matched modulation and detection chain, and the exact role of the channel classes \textit{Depolarizing Channel}, \textit{Dephasing Channel}, \textit{Erasure Channel}, \textit{Bosonic Channel}, \textit{Turbulence Channel}, and \textit{PMD Channel}. We also explain the current visualization layer, which projects received states onto constellation and Bloch representations for qualitative inspection. To keep the implementation-faithful,we provide a summary of the baseline execution, which uses a square 16-QAM embedding, a pretty-good-measurement detector constructed from the same reference-state codebook, and BER/SER. Finally, we position the channel-comparison as an entry point for broader future work, including equalization, quantum autoencoder, learning-based, and system-level algorithm integration.
\end{abstract}


\section{Introduction}
End-to-end communication modeling increasingly combines signal-processing abstractions, probabilistic channel models, and learned transmitter/receiver components \cite{oshea2017,dorner2018}. In the quantum setting, this integration is even more delicate because physically distinct impairments are often represented in different mathematical state spaces: qubit depolarization is naturally written as a completely positive trace-preserving map in a two-dimensional Hilbert space, while optical attenuation and turbulence are more naturally expressed in bosonic or wave-propagation form \cite{wilde2013,weedbrook2012}. A practical software benchmark therefore needs two properties at once. First, it must remain mathematically interpretable. Second, it must make heterogeneous channels executable within a common simulation loop.

QCommE2E addresses this requirement as an open-source software framework rather than a single benchmark script. The repository already spans channel models, transceiver abstractions, metrics, visualization utilities, end-to-end training hooks, equalization modules, and experiment notebooks. In other words, its scientific value is not limited to one comparison figure; it lies in providing a reusable environment in which end-to-end quantum communication workflows can be assembled, inspected, and extended. That broader role is useful for experiments, but it can be difficult for a new reader to reconstruct exactly which equations are being implemented, how dimensions are handled, which subsystem is responsible for each stage of the pipeline, and what the plotted constellations and Bloch points actually represent.

This paper fills that gap. Its contributions are fourfold. First, it presents QCommE2E as a tutorial-oriented open-source simulator for end-to-end quantum communication systems. Second, it maps the current source code to the governing equations for the active channel classes. Third, it explains the matched benchmark methodology used in the channel-comparison notebook, which serves here as the main use case for understanding the framework. Fourth, it summarizes the latest demonstrative results and clarifies what those numbers do not mean physically. The emphasis is therefore tutorial rather than novel: the manuscript documents the repository state as reflected by the current source files, tests, and generated notebook outputs, while showing how the same infrastructure could host more advanced algorithms later.

\section{Background and Preliminaries}
\subsection{Quantum states and channel interface}
QCommE2E adopts the density-matrix viewpoint. A quantum state is represented by a Hermitian positive semidefinite matrix $\rho$ with $\Tr(\rho)=1$, and a channel is represented as a completely positive trace-preserving map \cite{choi1975,kraus1983,wilde2013}
\begin{equation}
\E(\rho)=\sum_k E_k \rho E_k^\dagger,
\label{eq:kraus}
\end{equation}
where the Kraus operators satisfy $\sum_k E_k^\dagger E_k = I$. In the repository, this abstraction appears as the base class \texttt{BaseChannel}, whose sole required public method is
\begin{equation}
\rho_{\mathrm{out}} = \texttt{channel.apply}(\rho_{\mathrm{in}}).
\end{equation}
This single-state API is important: the current benchmark is vectorized only outside the channel, by iterating over transmitted states.

\subsection{Matched modulation and detection}
The benchmark uses a matched transmit/receive design so that BER and SER reflect channel effects rather than detector mismatch. For the current classes, the transmitter maps a classical symbol $s$ to a density matrix $\rho_s$, and the receiver applies a pretty-good measurement (PGM), also known as the square-root measurement \cite{hausladen1994,hausladen1996}. Given reference states $\{\rho_i\}$ and prior probabilities $\{p_i\}$, the detector forms
\begin{equation}
\bar{\rho} = \sum_i p_i \rho_i,
\end{equation}
and the Positive Operator-Valued Measure (POVM) elements
\begin{equation}
E_i = p_i\,\bar{\rho}^{-1/2}\rho_i\bar{\rho}^{-1/2}.
\label{eq:pgm}
\end{equation}
The hard decision is then
\begin{equation}
\hat{s} = \arg\max_i\, \Tr(E_i\rho_{\mathrm{out}}).
\label{eq:harddecision}
\end{equation}
This is exactly the logic implemented in \texttt{PrettyGoodMeasurementDetector}.

\subsection{Current modulation layers}
Two modulators are presently relevant to qubit-oriented baseline, the first is the \texttt{QPSKModulator} , which uses the codebook
\begin{align}
|\psi_0\rangle &= |0\rangle,
& |\psi_1\rangle &= |1\rangle, \\
|\psi_2\rangle &= \frac{|0\rangle+|1\rangle}{\sqrt{2}},
& |\psi_3\rangle &= \frac{|0\rangle-|1\rangle}{\sqrt{2}},
\end{align}
with density matrices $\rho_i = |\psi_i\rangle\langle\psi_i|$. Second, the active modulator is \texttt{QAMModulator} with $M=16$ which first constructs a square Gray-labeled classical QAM constellation \cite{proakis2008}, then embeds each complex point $\alpha\in\mathbb{C}$ into a qubit pure state by
\begin{equation}
|\psi(\alpha)\rangle = \frac{1}{\sqrt{1+|\alpha|^2}}
\begin{bmatrix}
1\\
\alpha
\end{bmatrix},
\qquad
\rho(\alpha)=|\psi(\alpha)\rangle\langle\psi(\alpha)|.
\label{eq:qamembedding}
\end{equation}
This embedding is an engineering device for running a matched detector on a qubit state space. It should not be interpreted as a complete optical coherent-state transmitter.

\subsection{Why channel comparison is the tutorial entry point}
Among the repository's available capabilities, channel comparison is presently the clearest tutorial use case because it exercises the full end-to-end path while remaining scientifically interpretable. A user can inspect the modulator, transmitter, channel, detector, metrics, and visualization outputs in one workflow without needing to introduce optimization or training loops. At the same time, this use case exposes the abstractions that future extensions would reuse. The same transmitter/receiver skeleton can later host adaptive equalizers, quantum autoencoders, reinforcement-learning agents, diffusion-based components, or more elaborate protocol layers, provided they honor the surrounding state and interface conventions.

\section{Methodology}
\subsection{Benchmark flow}
The quantum channel comparison notebook builds the chain in four steps. First, \texttt{build\_chain} instantiates a modulator and a PGM detector from the same reference-state codebook. Second, the notebook draws symbols uniformly from \texttt{mod.symbol\_alphabet()}. Third, the transmitter produces quantum states, and each state is passed through the selected channel. Fourth, the receiver detects a symbol label and computes BER and SER via
\begin{equation}
\mathrm{BER} = \frac{1}{N_b}\sum_{n=1}^{N_b} \mathbf{1}\{b_n\neq \hat{b}_n\},
\qquad
\mathrm{SER} = \frac{1}{N_s}\sum_{n=1}^{N_s} \mathbf{1}\{s_n\neq \hat{s}_n\}.
\label{eq:berser}
\end{equation}
In code, these are implemented by \texttt{compute\_ber} and \texttt{compute\_ser} as simple normalized Hamming mismatches.

The notebook also stores the transmitted states, received states, labels, and generated figures. The visualization path projects larger states onto their leading $2\times 2$ qubit block before generating constellation and Bloch plots. This is essential for the erasure and bosonic-surrogate cases, whose internal dimensions need not always remain equal to the logical detector dimension.

\subsection{Mathematical models}
The present source code implements the following equations.

\subsubsection{Depolarizing channel}
\texttt{DepolarizingChannel} applies the standard qubit depolarizing map \cite{wilde2013}
\begin{equation}
\E_{\mathrm{dep}}(\rho) = (1-p)\rho + p\,\frac{I_2}{2}.
\label{eq:depolarizing}
\end{equation}
This isotropically contracts the Bloch vectors and preserves the two-dimensional logical space.

\subsubsection{Dephasing channel}
\texttt{DephasingChannel} is implemented as phase damping rather than a Pauli-$Z$ flip mixture:
\begin{equation}
\rho = \begin{bmatrix}
\rho_{00} & \rho_{01}\\
\rho_{10} & \rho_{11}
\end{bmatrix}
\mapsto
\begin{bmatrix}
\rho_{00} & (1-p)\rho_{01}\\
(1-p)\rho_{10} & \rho_{11}
\end{bmatrix}.
\label{eq:dephasing}
\end{equation}
The populations remain unchanged, but coherence decays \cite{wilde2013}.

\subsubsection{Erasure channel}
\texttt{ErasureChannel} expands the Hilbert space by one orthogonal flag dimension. For an input of dimension $d$,
\begin{equation}
\E_{\mathrm{era}}(\rho) = (1-p)\rho \oplus p|e\rangle\langle e|,
\label{eq:erasure}
\end{equation}
where $|e\rangle$ is orthogonal to the original code space \cite{bennett1997}. In the detector, the original POVM is embedded into the larger space and an additional erasure element is appended with label $-1$.

\subsubsection{Bosonic thermal-loss channel}
\texttt{BosonicChannel} uses a truncated Fock-space Stinespring model. If $\rho_{\mathrm{th}}(n_{\mathrm{th}})$ is a thermal environment state and $U_\eta$ is a beamsplitter of transmissivity $\eta$, then
\begin{equation}
\E_{\mathrm{th}}(\rho) = \Tr_E\!\left[U_\eta\left(\rho\otimes \rho_{\mathrm{th}}\right)U_\eta^\dagger\right],
\qquad
\eta = 10^{-L/10},
\label{eq:bosonic}
\end{equation}
where $L$ is the loss in dB \cite{holevo1999,weedbrook2012}. In the current notebook sweep, \texttt{dim=2}, so the bosonic model is reduced to the vacuum/one-photon subspace.

\subsubsection{Atmospheric turbulence surrogate}
\texttt{TurbulenceChannel} samples a random transmissivity and then applies the repository's pure-loss map. The current implementation computes a pointing-loss factor
\begin{equation}
\eta_p = \exp\!\left(-2\left(\frac{\sigma_p}{w_0}\right)^2\right),
\label{eq:pointing}
\end{equation}
and combines it with a Gamma-Gamma or Malaga-inspired scintillation term \cite{alhabash2001}. The resulting transmissivity is clipped to $[0,1]$ and used in the pure-loss channel. This is therefore a fading surrogate, not a full phase-screen propagation model.

\subsubsection{PMD surrogate}
\texttt{PMDChannel} models polarization mode dispersion as repeated random principal-state-basis rotations. At section $k$, the code draws a Haar unitary $U_k$ and applies basis-dependent coherence attenuation
\begin{equation}
\nu = \exp\!\left(-\frac{1}{2}(\sigma_\omega\tau_{\mathrm{sec}})^2\right),
\qquad
\tau_{\mathrm{sec}} = \frac{\tau_{\mathrm{DGD}}}{\sqrt{N_{\mathrm{sec}}}},
\label{eq:pmdvisibility}
\end{equation}
followed by
\begin{equation}
\rho_{k+1}=U_k\,\mathcal{D}_\nu\!\left(U_k^\dagger\rho_kU_k\right)U_k^\dagger,
\label{eq:pmdsection}
\end{equation}
where $\mathcal{D}_\nu$ leaves diagonal entries unchanged and multiplies off-diagonals by $\nu$ \cite{foschini1991,gordon2000,alnaseri2026fading}. The final state is symmetrized numerically to maintain Hermiticity.

\section{Implementation Details}
\subsection{Repository classes and responsibilities}
The implementation is intentionally modular.
\begin{itemize}
\item \texttt{BaseChannel} enforces a single-state density-matrix interface and probability validation.
\item \texttt{Transmitter} delegates symbol-to-state conversion to the active modulator.
\item \texttt{Receiver} optionally applies equalization/decoding and then uses a detector to recover classical labels.
\item \texttt{PrettyGoodMeasurementDetector} computes a PGM from the modulator's own reference states and priors.
\item The notebook function \texttt{run\_simulation} coordinates symbol sampling, channel application, detection, metric computation, and figure export.
\end{itemize}
This decomposition is one of the strengths of the current code: channel classes can vary without rewriting the metric or plotting code, and future algorithmic modules can be inserted without collapsing the overall structure. In particular, the repository layout already leaves room for equalizers, autoencoders, distributed-network components, and learning-based agents to be integrated as additional decision-making layers around the same end-to-end path.

\subsection{Dimension handling and erasure logic}
The most delicate implementation issue is the dimension mismatch. Most channels preserve the modulator dimension, but \texttt{ErasureChannel} returns a $(d+1)\times(d+1)$ matrix. The detector handles this by embedding the reference POVM into the larger space and appending an erasure operator
\begin{equation}
E_{\mathrm{era}} = I_{d+1} - \sum_i E_i^{\uparrow},
\end{equation}
where $E_i^{\uparrow}$ denotes the reference element padded into the higher-dimensional state space. If this erasure element wins under \eqref{eq:harddecision}, the detector returns label $-1$. The corresponding bit mapping fills the bit vector with $-1$ entries, which makes hard erasures count as full mismatches in the present BER implementation.

\subsection{Visualization layer}
The visualization performs two implementation-specific projections.
\begin{enumerate}
\item For constellations, the leading qubit block is converted into a two-dimensional in-phase/quadrature surrogate. For QAM states, the code reconstructs the approximate classical point from the ratio $\rho_{10}/\rho_{00}$ in the leading qubit block.
\item For Bloch plots, the same leading $2\times 2$ block is converted into Bloch vectors and rendered with a direct Matplotlib 3D scatter routine. 
\end{enumerate}
These choices make the notebook visually useful, but they are best understood as diagnostic projections rather than invariant physical observables.

\subsection{Built-in software checks}
The current tests provide implementation-level verification. The channel tests check, among other things, density preservation for bosonic and turbulence channels, single-photon decay under pure loss, coherence attenuation for dephasing, orthogonal-flag enlargement for erasure, and reduced purity at larger DGD for the PMD surrogate. The transceiver tests verify the QPSK and 16-QAM modulators, the bit-label mappings, and the erasure-label behavior of the PGM detector. These tests do not prove physical completeness, but they do confirm that the implemented equations are internally consistent with the intended benchmark abstractions.

\section{Demonstrative Results: Channel Model Comparison}
\subsection{Current notebook configuration}
Table~\ref{tab:config} summarizes the channel model comparison notebook.

\begin{table}[ht]
\caption{Latest stored comparison configuration.}
\label{tab:config}
\centering
\begin{tabular}{ll}
\toprule
Setting & Value \\
\midrule
Notebook & \texttt{quantum\_channel\_comparison} \\
Modulation & square 16-QAM embedding \\
$M$ & 16 \\
Number of symbols & 4000 \\
Random seed & 123 \\
Detector & pretty-good measurement \\
Metrics & hard-decision BER and SER \\
Qualitative outputs & constellation and Bloch plots \\
\bottomrule
\end{tabular}
\end{table}

\begin{figure}[t]
\centering
\includegraphics[width=0.99\linewidth]{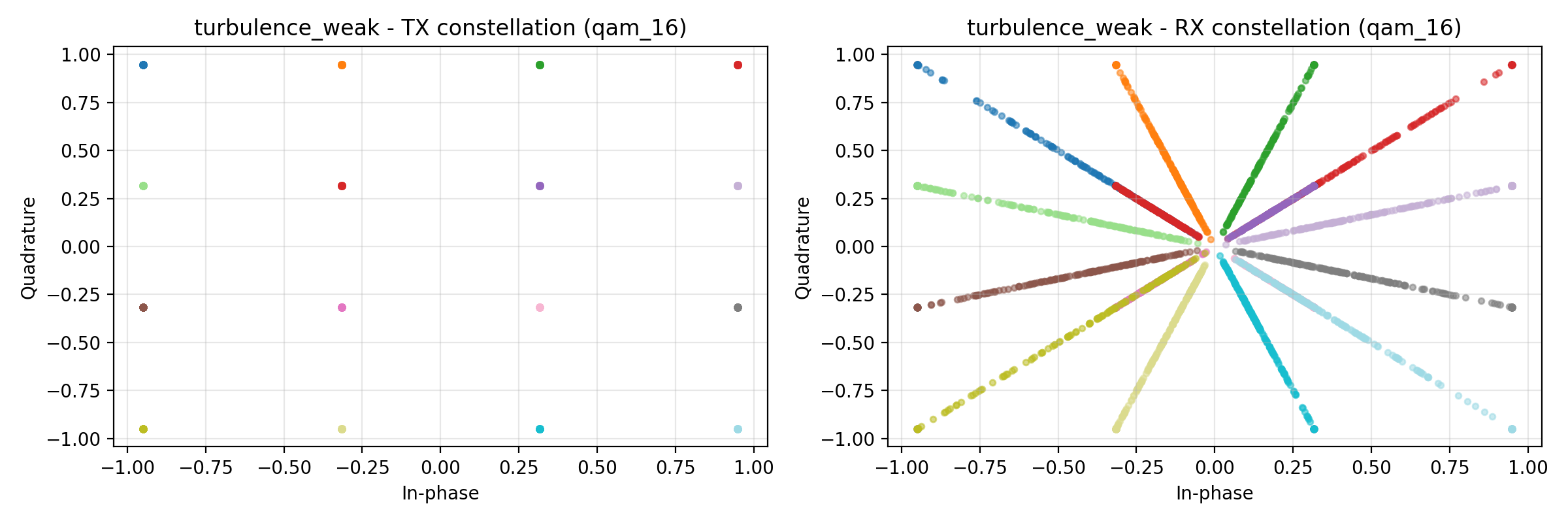}
\caption{Constellation diagram for the weak-turbulence configuration. Left: transmitted 16-QAM surrogate constellation. Right: corresponding received constellation.}
\label{fig:turbulence_constellation}
\end{figure}

\begin{figure}[t]
\centering
\includegraphics[width=0.99\linewidth]{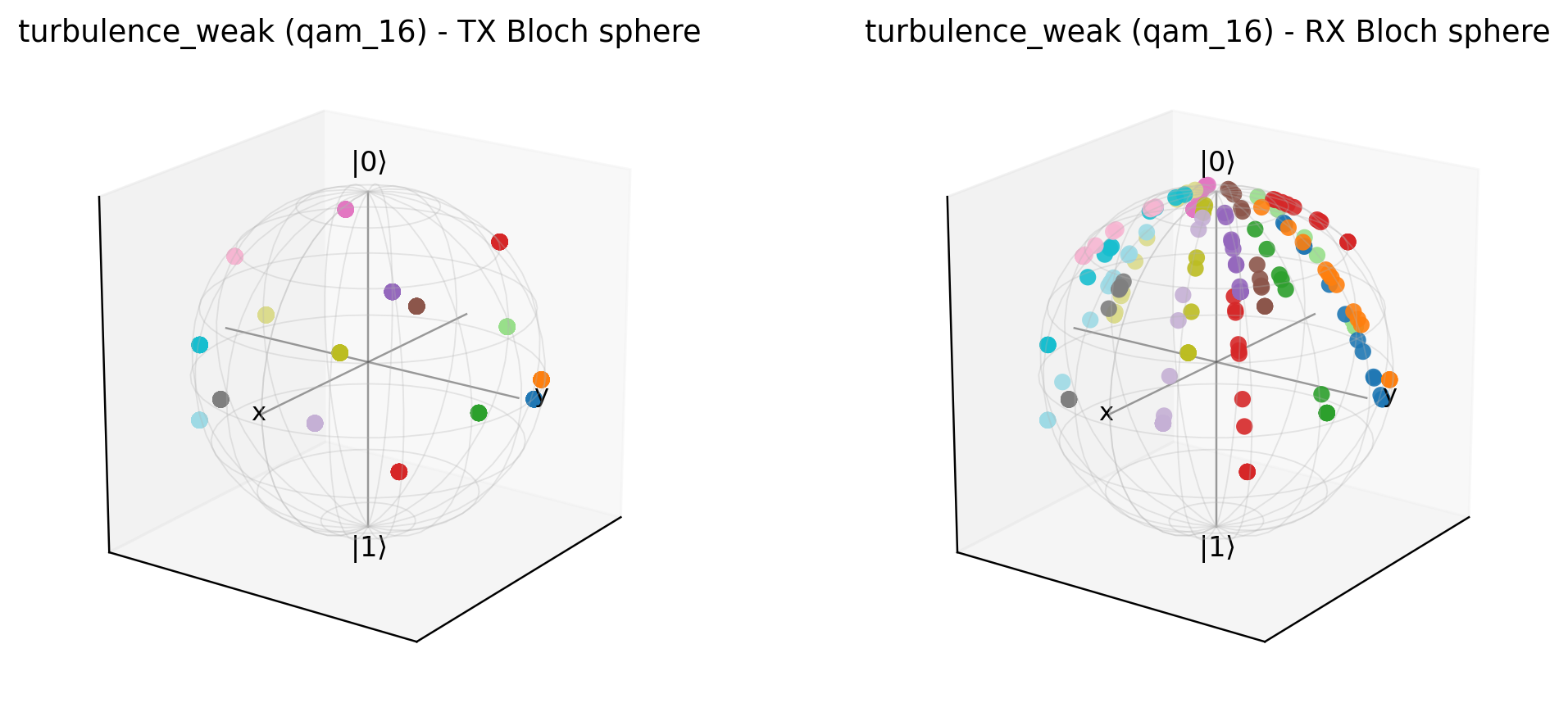}
\caption{Block spheres for the weak-turbulence configuration. Left: transmitted 16-QAM surrogate constellation. Right: corresponding received constellation.}
\label{fig:turbulence_bloch}
\end{figure}

\begin{figure}[t]
\centering
\includegraphics[width=0.99\linewidth]{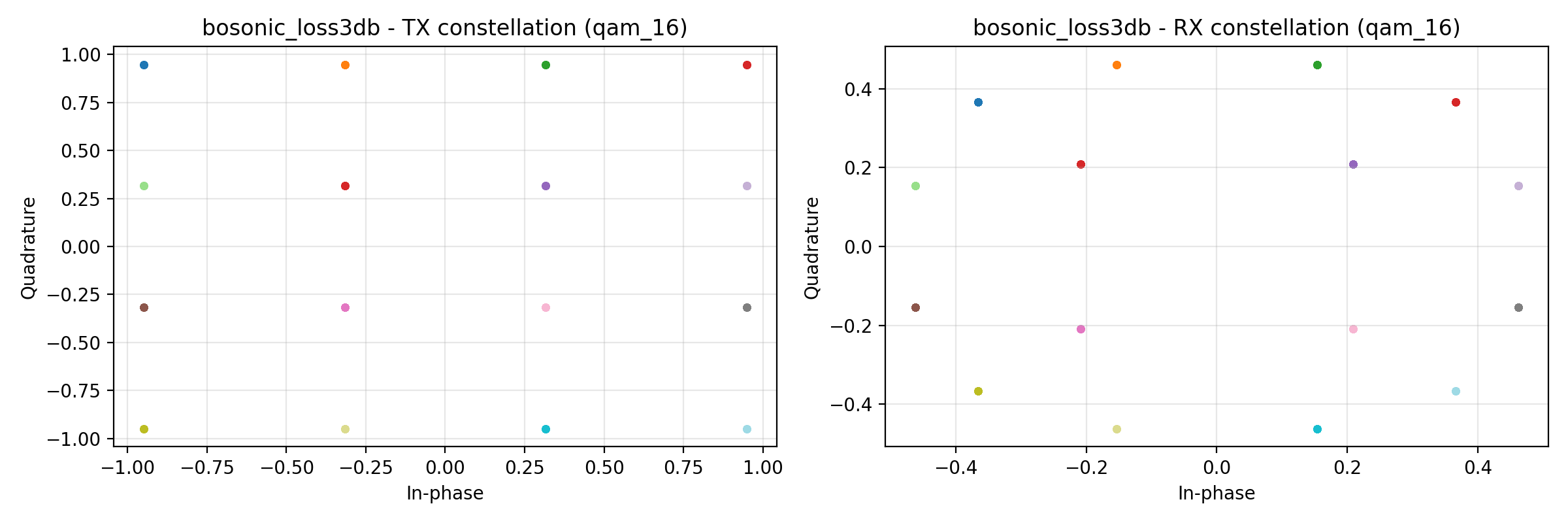}
\caption{Constellation diagram for the bosonic channel with 3dB loss configuration. Left: transmitted 16-QAM surrogate constellation. Right: corresponding received constellation.}
\label{fig:bosonic_constellation}
\end{figure}

\begin{figure}[t]
\centering
\includegraphics[width=0.99\linewidth]{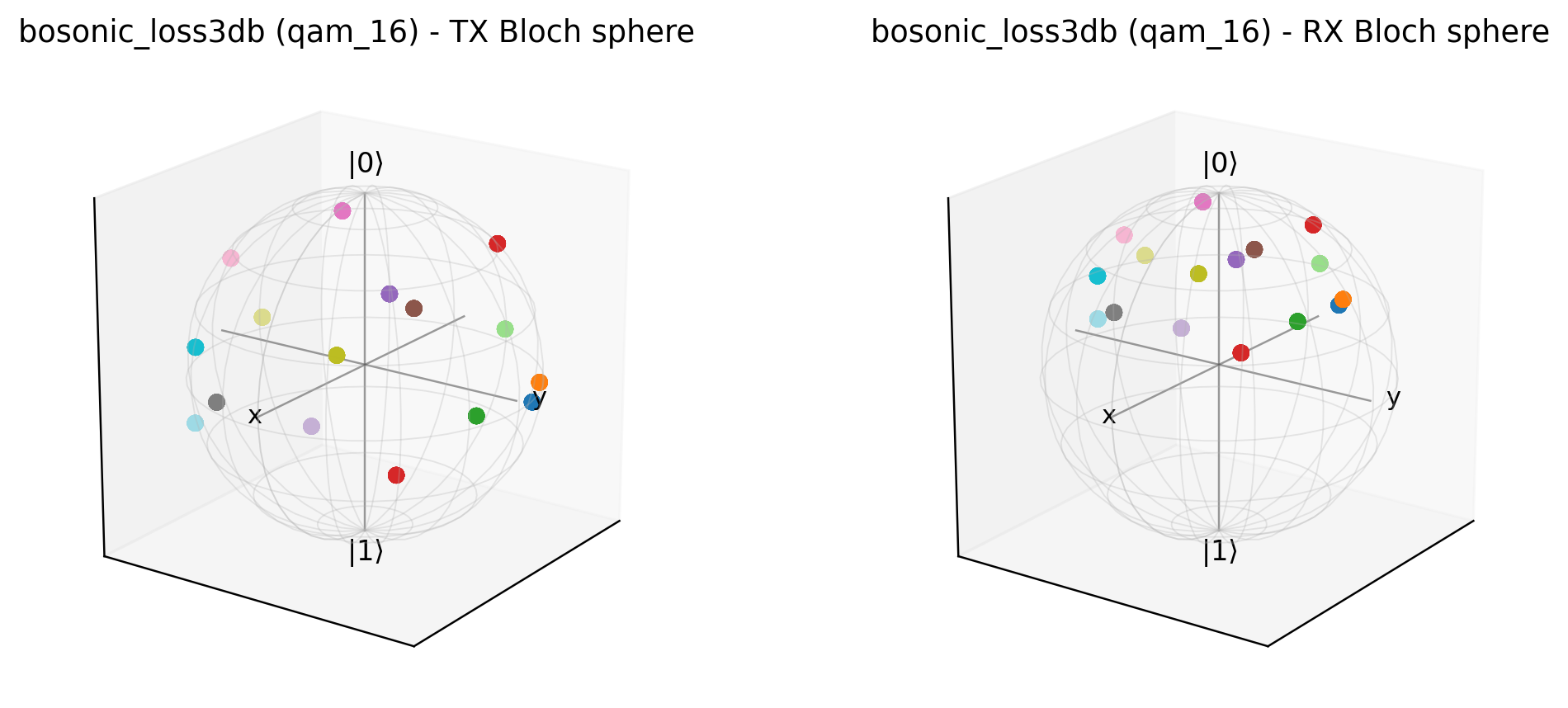}
\caption{Block spheres for the bosonic channel with 3dB loss configuration. Left: transmitted 16-QAM surrogate constellation. Right: corresponding received constellation.}
\label{fig:bosonic_bloch}
\end{figure}

Figures.~\ref{fig:turbulence_constellation}, ~\ref{fig:turbulence_bloch}, ~\ref{fig:bosonic_constellation}, and \ref{fig:bosonic_bloch} show one of the current color-coded outputs generated by the notebook. Fig.\ref{fig:bosonic_constellation} illustrates how the received qubit-block projection spreads the 16-QAM surrogate constellation under weak turbulence, while Fig.~\ref{fig:turbulence_bloch} shows the corresponding Bloch-space displacement of the same labeled states. Likewise, Fig.~\ref{fig:bosonic_constellation}, and Fig.~\ref{fig:bosonic_bloch} show the impact of bosonic channel model under 3dB loss. 

In practice, these qualitative plots are valuable because they expose effects that the BER/SER table alone cannot. For example, the turbulence and bosonic models can be seen to relocate symbol clusters gradually, whereas the erasure model removes probability mass from the logical subspace altogether. That distinction is important when interpreting why two channels can both degrade accuracy while doing so through very different mechanisms.

\section{Discussion and Limitations}
The current framework is useful, but it should not be oversold. Its main limitations are as follows.

First, the benchmark mixes models of different physical fidelity. The depolarizing, dephasing, and erasure channels are canonical finite-dimensional quantum-information models, whereas the bosonic, turbulence, and PMD classes are reduced optical surrogates built to fit a common density-matrix pipeline. This is a pragmatic software choice, not a claim that the channels are physically interchangeable.

Second, the comparison is currently memoryless across symbols. Even the PMD and turbulence classes, which are motivated by propagation phenomena with temporal structure in real systems, are applied one state at a time in the notebook. As a result, burst effects, temporal fading correlation, and adaptive equalization dynamics are absent from the specific comparison documented here.

Third, the current optical surrogates are low-dimensional. In particular, the bosonic notebook settings use \texttt{dim=2}, so the model is restricted to the vacuum/one-photon subspace. This is sufficient for a software benchmark, but it is not a full continuous-variable optical simulation \cite{weedbrook2012}.

Fourth, the results are hard-decision only. The BER/SER path uses the final detector label, not soft posterior information. Consequently, the present tables are not information-theoretic characterizations of channel capacity; they are engineering summaries of how the current codebook/detector pair behaves at the tested settings.

Fifth, the latest stored results are based on a single run with 4000 symbols. That is adequate for a tutorial notebook and for visual inspection, but not for publication-grade BER curves or statistically tight confidence intervals.

Finally, the current QAM embedding in \eqref{eq:qamembedding} is a repository design choice. It is a clean way to keep the benchmark within a qubit detector pipeline, but it is not a complete physical model of a quantum-optical QAM transmitter.

These limitations do not reduce the value of QCommE2E as open-source infrastructure. Instead, they define the current frontier of the tutorial implementation. The repository is already well positioned for incremental extensions such as soft-output detectors, temporal channel correlation, richer bosonic truncations, adaptive equalization, trainable encoders/decoders, and learning-based control policies. The point of the present paper is therefore twofold: to explain the current end-to-end simulator faithfully, and to make clear where additional algorithms can attach later without changing the conceptual backbone.

\section{Conclusion}
This paper presented QCommE2E as an open-source, tutorial-oriented simulation environment for end-to-end quantum communication systems. The unifying idea is simple: modular transmitters, receivers, channels, metrics, and visualization routines are organized around a common density-matrix workflow, and the current channel-comparison notebook provides the most direct executable demonstration of that design. Within that framework, QCommE2E currently supports textbook qubit channels, a flagged erasure channel, a truncated bosonic thermal-loss model, a turbulence-induced fading surrogate, and a spectrum-averaged PMD surrogate.

The main practical lesson is that the repository already contains a coherent tutorial simulator, but it must be interpreted at the level of its present abstractions. Some models are canonical, some are reduced surrogates, and the current notebook results are demonstrative rather than final scientific claims. Precisely because of that, the framework is useful: it gives a clear common language for extending QCommE2E toward richer optical dimensions, time-correlated channels, soft-information metrics, adaptive equalization, trainable end-to-end algorithms, and tighter experiment protocols. In that sense, the quantum channel comparison workflow is not the end of the story; it is the primary worked example through which the broader simulator can be understood and extended.

\section*{Code Availability}

The implementation of QCommE2E is available at:

\url{https://github.com/oalnaseri/QCommE2E}






\bibliographystyle{IEEEtran}
\bibliography{references}

\end{document}